\documentstyle[twoside]{article}

\catcode`\@=11
\long\def\@makefntext#1{
\protect\noindent \hbox to 3.2pt {\hskip-.9pt
$^{{\eightrm\@thefnmark}}$\hfil}#1\hfill}               %CAN BE USED

\def\@makefnmark{\hbox to 0pt{$^{\@thefnmark}$\hss}}    %ORIGINAL

\def\ps@myheadings{\let\@mkboth\@gobbletwo
\def\@oddhead{\hbox{}
\rightmark\hfil\eightrm\thepage}
\def\@oddfoot{}\def\@evenhead{\eightrm\thepage\hfil
\leftmark\hbox{}}\def\@evenfoot{}
\def\sectionmark##1{}\def\subsectionmark##1{}}

\oddsidemargin=\evensidemargin
\newcounter{sectionc}\newcounter{subsectionc}\newcounter{subsubsectionc}

\renewcommand{\section}[1] {\vspace{12pt}\addtocounter{sectionc}{1}
\setcounter{subsectionc}{0}\setcounter{subsubsectionc}{0}\noindent
        {\tenbf\thesectionc. #1}\par\vspace{5pt}}
\renewcommand{\subsection}[1]
{\vspace{12pt}\addtocounter{subsectionc}{1}
      \setcounter{subsubsectionc}{0}\noindent
      {\bf\thesectionc.\thesubsectionc.{\kern1pt \bfit
#1}}\par\vspace{5pt}}
\renewcommand{\subsubsection}[1]
      {\vspace{12pt}\addtocounter{subsubsectionc}{1}
      \noindent{\tenrm\thesectionc.\thesubsectionc.\thesubsubsectionc.
      {\kern1pt \tenit #1}}\par\vspace{5pt}}
\newcommand{\nonumsection}[1] {\vspace{12pt}\noindent{\tenbf #1}
        \par\vspace{5pt}}

\newcounter{appendixc}
\newcounter{subappendixc}[appendixc]
\newcounter{subsubappendixc}[subappendixc]
\renewcommand{\thesubappendixc}{\Alph{appendixc}.\arabic{subappendixc}}
\renewcommand{\thesubsubappendixc}%
{\Alph{appendixc}.\arabic{subappendixc}.\arabic{subsubappendixc}}

\renewcommand{\appendix}[1] {\vspace{12pt}
        \refstepcounter{appendixc}
        \setcounter{figure}{0}
        \setcounter{table}{0}
        \setcounter{lemma}{0}
        \setcounter{theorem}{0}
        \setcounter{corollary}{0}
        \setcounter{definition}{0}
        \setcounter{equation}{0}
        \renewcommand{\thefigure}{\Alph{appendixc}.\arabic{figure}}
        \renewcommand{\thetable}{\Alph{appendixc}.\arabic{table}}
        \renewcommand{\theappendixc}{\Alph{appendixc}}
        \renewcommand{\thelemma}{\Alph{appendixc}.\arabic{lemma}}
        \renewcommand{\thetheorem}{\Alph{appendixc}.\arabic{theorem}}

\renewcommand{\thedefinition}{\Alph{appendixc}.\arabic{definition}}

\renewcommand{\thecorollary}{\Alph{appendixc}.\arabic{corollary}}
        \renewcommand{\theequation}{\Alph{appendixc}.\arabic{equation}}
        \noindent{\tenbf Appendix \theappendixc #1}\par\vspace{5pt}}
\newcommand{\subappendix}[1] {\vspace{12pt}
        \refstepcounter{subappendixc}
        \noindent{\bf Appendix \thesubappendixc. {\kern1pt \bfit #1}}
        \par\vspace{5pt}}
\newcommand{\subsubappendix}[1] {\vspace{12pt}
        \refstepcounter{subsubappendixc}
        \noindent{\rm Appendix \thesubsubappendixc. {\kern1pt \tenit #1}}
        \par\vspace{5pt}}

\newcommand{\smalllineskip}{\baselineskip=10pt}

\def\eightcirc{
\begin{picture}(0,0)
\put(4.4,1.8){\circle{6.5}}
\end{picture}}
\def\eightcopyright{\eightcirc\kern2.7pt\hbox{\eightrm c}}

\def\abstracts#1#2#3{{

\centering{\begin{minipage}{4.5in}\baselineskip=10pt\footnotesize
        \parindent=0pt #1\par
        \parindent=15pt #2\par
        \parindent=15pt #3
        \end{minipage}}\par}}

\renewenvironment{thebibliography}[1]
        {\frenchspacing
         \ninerm\baselineskip=11pt
         \begin{list}{\arabic{enumi}.}
        {\usecounter{enumi}\setlength{\parsep}{0pt}
         \setlength{\leftmargin 12.7pt}{\rightmargin 0pt} %FOR 1--9 ITEMS
         \setlength{\itemsep}{0pt} \settowidth
        {\labelwidth}{#1.}\sloppy}}{\end{list}}

\newcounter{itemlistc}
\newcounter{romanlistc}
\newcounter{alphlistc}
\newcounter{arabiclistc}

\newcommand{\fcaption}[1]{
        \refstepcounter{figure}
        \setbox\@tempboxa = \hbox{\footnotesize Fig.~\thefigure. #1}
        \ifdim \wd\@tempboxa > 5in
           {\begin{center}
        \parbox{5in}{\footnotesize\smalllineskip Fig.~\thefigure. #1}
            \end{center}}
        \else
             {\begin{center}
             {\footnotesize Fig.~\thefigure. #1}
              \end{center}}
        \fi}

\newcommand{\tcaption}[1]{
        \refstepcounter{table}
        \setbox\@tempboxa = \hbox{\footnotesize Table~\thetable. #1}
        \ifdim \wd\@tempboxa > 5in
           {\begin{center}
        \parbox{5in}{\footnotesize\smalllineskip Table~\thetable. #1}
            \end{center}}
        \else
             {\begin{center}
             {\footnotesize Table~\thetable. #1}
              \end{center}}
        \fi}

\def\@citex[#1]#2{\if@filesw\immediate\write\@auxout
        {\string\citation{#2}}\fi
\def\@citea{}\@cite{\@for\@citeb:=#2\do
        {\@citea\def\@citea{,}\@ifundefined
        {b@\@citeb}{{\bf ?}\@warning
        {Citation `\@citeb' on page \thepage \space undefined}}
        {\csname b@\@citeb\endcsname}}}{#1}}

\newif\if@cghi
\def\refcite{\@cghitrue\@ifnextchar [{\@tempswatrue
        \@citex}{\@tempswafalse\@citex[]}}
\def\refcitelow{\@cghifalse\@ifnextchar [{\@tempswatrue
        \@citex}{\@tempswafalse\@citex[]}}
\def\@cite#1#2{{$\null^{#1}$\if@tempswa\typeout
        {IJCGA warning: optional citation argument
        ignored: `#2'} \fi}}

\def\@refcitex[#1]#2{\if@filesw\immediate\write\@auxout
        {\string\citation{#2}}\fi
\def\@citea{}\@refcite{\@for\@citeb:=#2\do
        {\@citea\def\@citea{, }\@ifundefined
        {b@\@citeb}{{\bf ?}\@warning
        {Citation `\@citeb' on page \thepage \space undefined}}
        \hbox{\csname b@\@citeb\endcsname}}}{#1}}

\def\@refcite#1#2{{#1\if@tempswa\typeout
        {IJCGA warning: optional citation argument
        ignored: `#2'} \fi}}

\def\refcite{\@ifnextchar[{\@tempswatrue
        \@refcitex}{\@tempswafalse\@refcitex[]}}

\def\pmb#1{\setbox0=\hbox{#1}
        \kern-.025em\copy0\kern-\wd0
        \kern.05em\copy0\kern-\wd0
        \kern-.025em\raise.0433em\box0}

\def\fnt#1#2{\footnotetext{\kern-.3em
        {$^{\mbox{\scriptsize #1}}$}{#2}}}

\def\fpage#1{\begingroup
\voffset=.3in
\thispagestyle{empty}\begin{table}[b]\centerline{\footnotesize #1}
        \end{table}\endgroup}

\def\runninghead#1#2{\pagestyle{myheadings}
\markboth{{\protect\footnotesize\it{\quad #1}}\hfill}
{\hfill{\protect\footnotesize\it{#2\quad}}}}
\font\tenrm=cmr10
\font\tenit=cmti10
\font\tenbf=cmbx10
\font\bfit=cmbxti10 at 10pt
\font\ninerm=cmr9

\font\eightrm=cmr8

\def\qed{\hbox{${\vcenter{\vbox{                      %HOLLOW SQUARE
   \hrule height 0.4pt\hbox{\vrule width 0.4pt height 6pt
   \kern5pt\vrule width 0.4pt}\hrule height 0.4pt}}}$}}

\begin{document}

\runninghead{George Svetlichny}
{Long Range Correlations and Relativity $\ldots$}

%\normalsize\textlineskip
\thispagestyle{empty}\setcounter{page}{1}
\vspace*{0.88truein}
\fpage{1}

\centerline{\bf LONG RANGE CORRELATIONS AND RELATIVITY:}
\centerline{\bf METATHEORETIC CONSIDERATIONS}
\vspace*{0.035truein}

\vspace*{0.37truein}
\centerline{\footnotesize George Svetlichny}

\centerline{\footnotesize \it
Departam\'ento de Matem\'atica, Pontif\'{\i}cia Universidade Cat\'olica}
\baselineskip=10pt
\centerline{\footnotesize \it
Rua Marqu\^es de S\~ao Vicente 225}
\baselineskip=10pt
\centerline{\footnotesize \it
22453-900 G\'avea, Rio de Janeiro RJ, Brazil}
\baselineskip=10pt
\centerline{\footnotesize \it
e-mail: svetlich@mat.puc-rio.br}

%\date{\today}

\baselineskip 5mm

\vspace*{0.21truein}

\abstracts{Action-at-a-distance is a generic  property of physical
theories. As such, it is not a fruitful idea in theory building.
Its absence confers a rigidity on a theory which we exemplify
through analysis of long-range correlations of the EPR-type in 
relativistic theories. Rigidity is desirable for fundamental theories,
and theory building should focus on structurally unstable properties,
making action-at-a-distance a side issue. 
Though apparent superluminal effects seem to be 
present in many present-day physical theories, 
we maintain that they are not a basis for action-at-a-distance. 
}{}{}

%\pacs{PACS numbers: 03.50.-z, 03.50.De}

\bigskip

$$$$

\section{Introduction}

Ever since the famous EPR debate [\refcite{einst,bohr}], 
there has been widespread speculations
that long range quantum correlations are indication of some sort of
action-at-a-distance. The situation is extremely subtle and the presence
or absence of action-at-a-distance depends a lot on philosophical
concerns [\refcite{desp}], though many physicist would deny this. 
We do not want to
enter this debate. The purpose of this essay is to expound the
consequences of the existence of these correlations, in the peculiar way
that they manifest themselves, concerning possible physical theories.
Whether these theories are to be of an action-at-a-distance type or not, 
is something
that depends on the details, interpretations, and conceptual 
content of the theories. The
information we shall convey is to a large extent independent of 
all this, and so must
be heeded by all interested in this debate.

\section{Metatheory of Superluminal Communication}
Although the light cone has been considered  an impenetrable barrier,
much of  present-day physics, based on this impenetrability, actually 
predicts a variety of phenomena that
seem to bridge the gap between the subluminal and superluminal.
Even plain classical Maxwell electrodynamics has superluminal 
solutions [\refcite{rod}].
Quantum gravity effects allow for light to propagate outside the
gravitational ``light cone" calling into question just what is the exact
causal structure of space-time [\refcite{konst}].
Extremely general quantum field theoretic 
considerations seem to imply superluminal influences [\refcite{heger}].
These examples can be multiplied many-fold. 
Effects seem to slip across the light cone in spite of a firm
theoretic resolve to contain them. This fact alone is remarkable given
the wide variety of theories for which this happens, and some general
characteristic, unsuspected up to now, 
could probably be discovered to explain it.
In any case, one need not be surprised that a
widespread debate continues. A good portion of this debate seems to 
exist on the fringe of
main-stream physics, in obscure journals and on the internet,
and one should ask why.

What is needed of course is some way of proceeding without entering the
details of particular theories. Superluminal situations arise out of
details. What is not clear is their effect on life  as we live it. Is
it action-at-a-distance? This depends on many conceptual subtleties. Can one send a 
message?
This is more straightforward. Can I hold a conversation with my relative
on a Mars colony with each remark followed immediately by a 
response as in a normal
living-room conversation, not suffering the usual speed-of-light 
time delay? Will we both agree, upon meeting each other 
two years later,
 that each one indeed said what the other remembered? 
These are
uncontestable gross effects. This is much like machinery,
wheels turning, lights flashing, and bar-room shouting. 
What one needs to establish
is whether  the superluminal effects implicit in present-day 
main-stream physics can have this ``gross thermodynamic quality" 
or are they
just peculiar
properties of the theoretical apparatus which do not lead to the gross
effects that are needed. The firm belief, almost faith, on part of the
majority of ``working physicists" that the second alternative is the
truth,
contributes to keeping a portion of this debate mostly on the fringe. 
The situation has not been
helped by an abundance of arguments based on a faulty or partial
understanding of modern physics and that are then easily if not
trivially put down, often accompanied by vehement arguments 
by the proponent
that this was not done. In a way this is reminiscent of the 
abundance in centuries past of
proposals for perpetual motion, before a true
understanding of thermodynamics was achieved.

Nick
Herbert [\refcite{herb}] for instance 
proposed an arrangement that would duplicate a photon state through
stimulated emission. Superluminal
communication is then easily achieved. 
This proposal quickly provoked various
rebuttals 
[\refcite{dieks,milo,woot}]
to the effect that no linear state transformer can clone a
quantum state, an instance of the so called ``No Cloning Theorem". 
Stimulated emission simply does not work the way
the proponent assumed it does, an assumption that implied 
non-linearities in quantum mechanics of a type never observed.

The argument that in the EPR situation no superluminal 
communication is possible 
[\refcite{eb2,eb3,ghio,ghim}] is that 
the statistical behavior of any
detector placed on one arm of an EPR apparatus is completely
independent of what is done on the other arm. 
Thus even though action-at-a-distance{\ }
may be present as  part of the ``hidden gears of nature" that do
indeed 
spin on the microscopic scale, our access to these processes, 
it is claimed,
 is such
that we cannot create gross quality effects. Some dispute this
last claim, but none have yet built a working device.
In what way  should the EPR situation be 
considered action-at-a-distance even though no energy or matter is 
transferred in
any way, is to be left to the philosophically inclined [\refcite{desp}]. 
We do not wish to
deny the value of philosophical analysis, but call attention to the fact
that the action-at-a-distance debate has forced upon us a level of 
philosophical subtlety usually absent in physical discourse. One does
not need to enter such realms as much can still be done with more
directly physical considerations.

Superluminal communication and action-at-a-distance  
are not logically equivalent,
but of course closely related. 
The hidden-gears-of-nature position shows that one could maintain
action-at-a-distance  without producing gross quality effects, but
discounting direct transfer of information to the receiver's mind
 (and only in certain theories of the mind at that), 
it's hard to see how one can 
receive a
distant signal without the sender ``acting" upon a physical entity that
manifests the signal. 
The scientific merit of a hidden-gears-of-nature posture is dubious,
and by action-at-a-distance  we shall mean a gross quality one, and so
we treat 
action-at-a-distance  and superluminal communication  as 
being about the same thing.

One cannot discount relativity. If we simply deny universality 
of special relativity 
then the debate becomes fruitless as no new guiding principle 
is brought forth to substitute the supremely powerful one that was 
discarded. As a local (space-time tangent-plane) symmetry, 
special relativity 
has been borne out with great precision. One cannot furthermore 
discount cosmology, as
gross quality superluminal effects would surely, one expects, 
have profound effects on the structure of the universe.

One could suppose that relativity is not universally valid, and that
there is a privileged frame, roughly, for instance, that of isotropic
background cosmic radiation or that of isotropic galactic red-shifts, 
which is sensed by some processes and with
respect to which instantaneous action-at-a-distance  is possible. 
Any such theory
would have no causality problems, would  alleviate some of the
paradoxical features of the abundant apparent superluminal effects,   
and could probably
be stretched to fit the known facts,
but conceptually and 
fundamentally would be very different from the current one. 
In the end a theory must be judged by experiment. Until then one should
inquire if the 
tension  resulting from stretching it to fit the facts is 
greater or not than the one in the current
theory.

To make some headway into this complicated business, one can, instead of
searching for theories that allow action-at-a-distance  and then 
stretching them to make them
hang on the known physical facts, take the inverse approach. Postulate a
principle that disallows gross quality action-at-a-distance  effects, 
something 
akin to the second law of thermodynamics, and see what this means for
possible physical theories. 

A useful notion is that of theory space. Consider all possible physical
theories, or to be less vague, all possible physical theories of a given
type. One can conceive of a type of topology on this space given by
proximity of predicted results. To dispel some of the aetherial
quality  of this, consider a large set of descriptions of experimental
arrangements, each with a finite number of possible outcomes. 
A theory is then a function that
associates to each description the probability distributions of the
outcomes. The set of all such functions is then the theory space, and
the topology may well be the weak topology, that is, one for which a
neighborhood is defined by proximity of the prediction of probability
of a finite number of events. This approach is very akin
to the ``empirical logic" approach pioneered by Foulis and Randall 
[\refcite{foul}]. 
Many  questions concerning the space of theories can be precisely
formulated and investigated in this manner.

Present day physical theory is a point in this space. It is surrounded by
theories that are proximate in experimental predictions but that may
well be radically different in other aspects.  One is interested mainly
in a weak neighborhood of the present-day theory as any serious alternative
theory must agree with well-established results predicted by the
present-day theory.
The relevant concept now is
that of structural stability. Do all neighboring theories share a
property of the present-day theory or not? If not, how big, in an
appropriate sense, is the set of neighboring theories that do? One has
to work with
some technical notion of ``almost all". Topologically this 
could be ``dense" or
``second category". We say a property is structurally unstable if almost
all theories in a neighborhood violate it, and structurally stable if
almost all share it. Structural instability, in spite of this 
weak-sounding designation, is a sign of a strong
fundamental theory as one then has sufficient reason to differentiate it
from neighboring ones.

What emerges from metatheoretic studies is
strong evidence toward the  claim  that, first of all, Lorentz
covariance, the existence of self-subsisting physical states, and the
existence of enough EPR-type long-range correlations, practically
characterize present-day linear quantum theory, and second, that the 
absence of gross quality superluminal effects
is  structurally
unstable, that is,  almost all theories in the neighborhood of present day
physics allow
superluminal communication. 
The first hypothesis  almost certainly can  be replaced by general
covariance of general relativity, the other two will be explained
in the course of this essay.
The structural instability explains many facts. One is a
sociological one. Why is there an abundance of proposals of superluminal
signaling devices based on present-day physics? First of all, the 
abundance of apparent superluminal effects in a variety of physical
theories, pointed out at the beginning, 
means that many will stumble upon at least one of them, 
and so be induced to believe that such a device may be possible. 
Second, the slightest
misunderstanding of present-day physics, or the slightest miscalculation
within it, places one at a neighboring theory and then it's practically
inevitable that one will conclude that superluminal communication is 
{\em really\/}
possible. The argument may seem watertight as the most subtlest of
errors will be enough to lead to the conclusion. Another fact is the great
robustness of present-day theories. Serious alternatives seem to run up
against insurmountable difficulties. Thus Weinberg's non-linear quantum
theory was abandoned by its creator exactly because he could not formulate
a relativistic version [\refcite{wein}]. 
One now understands why he failed. 

What do we mean by ``self-subsisting physical states"? In quantum
theory (in the Schr\"odinger picture), a 
physical state evolves deterministically by a unitary group in
Hilbert space. Such a state is generally created at some time and
destroyed later in a measurement process, but the deterministic
evolution can be extended to both temporal infinities. In particular it
can be extended to a time before its creation, which means it could have
been created at an earlier time and in some other place. Likewise the
evolution can be extended to a time after its destruction, which means
it could have been subjected to measurement at a later time and in
different place. The state is thus an autonomous physical entity having
no memory of its birth nor any prescience of its demise. Regardless of
the ontological status of such entities, physical theories use them as
algorithmic devices to compute joint probabilities of observed events.  A
sequence of events is then seen as the interaction of a state with a
measurement apparatus (or something akin to it)
by which the state is modified and then evolves until
the next interaction when it suffers another modification followed by
another evolution, and so on. 
Thus joint probabilities of events are computed using the
interpolating existence of evolving self-subsisting entities. 
This is not a logically necessary picture.  One can
take the strange sounding position that physical states are not really 
necessary to do physics, as one can conceive of ways of calculating joint
probabilities without the use of such interpolating entities. In fact
certain patterns of probabilities cannot be interpreted this 
way. The ``consistent histories" approach to quantum
mechanics [\refcite{hartle,omnes}] in fact abolishes to a large extent the 
reliance on 
self-subsisting physical states and can easily 
produce examples [\refcite{hartle}]
where  joint probabilities cannot be explained by such. Such an approach
also suggests [\refcite{svet:nlqm}] that the obstruction to relativistic
non-linear quantum mechanics, so lamented by Weinberg [\refcite{wein}],
can be overcome.

To begin our analysis [\refcite{svet:nh}], we work with
the hypothesis that there may be some physical processes that
do not conform to usual quantum mechanics
but that these only take place in 
very particular situations, whereas for the vast
majority of other situations,  such as experiments done  
up to now, any deviation from normal quantum mechanical predictions is
imperceptible.  One could thus posit a photon cloner
that acts in a non-linear fashion, and that it
can take part in an experiment in which normal
quantum mechanics is adequate for processes not involving it.
Explicitly one assumes that, in any given inertial frame, up
to the use of an unconventional device, the usual quantum mechanical
reasoning can be used, including the projection rule. Up to such a
moment, ordinary quantum mechanics determines what the physical state is.
At the point of using the unconventional 
device one of course must explicitly
say 
what would happen (a photon would be cloned in the above cited example).

What one succeeds in showing
under these hypotheses is that certain types of deviations, 
specifically non-linearities
and lack of true randomness of outcomes,  allow for superluminal signals.
This makes ordinary quantum
mechanics a structurally unstable theory in relation to the property of
not allowing superluminal communication.  This is important as many
proponents of modifications to ordinary quantum mechanics are in fact
implicitly assuming our hypotheses and so face a real risk of 
coming into conflict
with relativity, assuming the existence of   
superluminal communication is such a conflict.

More explicitly, [\refcite{svet:nh}] shows that, given our hypotheses, 
1) in a Hilbert space of dimension at least three, 
any state transformer, including temporal evolution,
must be given by a linear transformation
of density matrices, 
2) if a state transformer takes pure states into pure states and has
at least two states in its range, it can be implemented either by a
linear or an anti-linear operator, and 
3) randomness of possible outcomes in one experiment implies
randomness of outcomes in all.

It must be emphasized, as was mentioned before, 
that the above assumptions are about
formalism and not about interpretation.  What is postulated is 
an  altered
formalism associated to what is  generally  known as the Copenhagen
interpretation, but no interpretational hypotheses are made. State
collapse is used, but no assumption as to its ontological nature is made,
only that it's a legitimate calculating device for joint probabilities.
What the results say is that joint probabilities cannot
be calculated by certain rules if superluminal communication is to be ruled
out. 

It should also be noted that part of our understanding about the
standard formalism is that it's capable of giving account of a
relativistically covariant theory. This is not straightforwardly obvious
given the instantaneous nature of wave function
collapse [\refcite{aha1,aha2}], but this
does not preclude lorentz covariance of observable quantities. What the
standard formalism lacks is thus {\em manifest\/} covariance while being
able to provide for covariance of measurable magnitudes. It's precisely
this fact that makes the theory structurally unstable, for a
perturbation in the formalism is likely to make the manifest
non-covariance capable of producing real effects, such as superluminal
communication. In fact all theories that incorporate any frame-dependent 
notions, such as {\em temporal\/} evolution, and which have no gross
quality superluminal effect can probably be interpreted as a ``hidden
gears of nature" theory. A small change in the theory can expose the
hidden gears and make them accessible to our manipulation and so
superluminal communication becomes possible. A recipe for constructing
a superluminal
signaling device is generally very easy to discover in any such modified
theory, for instance, Gisin [\refcite{gisin}] has done it 
explicitly for Weinberg's non-linear quantum mechanics.

There is also an argument that  relates the absence of 
gross quality superluminal effects and the second law of
thermodynamics showing that under certain hypotheses, which include special
relativity, superluminal
communication can be used to foil the second law of thermodynamics. This
is because with superluminal communication, information can flow backward
in time. One can then foresee details of normal thermodynamic 
fluctuations and take advantage of them to extract work from heat.
This points out the thermodynamic character of any action-at-a-distance  
proposal within special relativity, a 
connection that was also pointed out by Elitzur [\refcite{elitzur}]. 

A striking feature of the above conclusions is their generality. This in
fact throws doubt on the emphasis given to superluminal communication 
and makes one suspect a more fundamental tension in alternate theories. 
In fact, the presence of superluminal signals 
as they emerge from the analysis,
{\em per se\/} already contradicts
relativity. Consider a superluminal signaling device
making use of the ``exposed" state-collapse mechanism and that is to 
operate
between two distant locations in the  reference frame of two
observers at relative rest. According to the general results,
if  the first observer invokes the signaling process, then the
second observer will, after a
negligible time interval, detect it. We can
say that for the second observer the {\em onset\/} of the signal is
practically simultaneous with the initiating event. Onset is a physical
event and so all observers ought to agree where in space-time it
occurred. Consider how the same situation is seen in a reference frame
of a moving observer. He would see a different initial state, find that
his physics is described   by  possibly different deviant equations,
but, assuming relativity, he does all his reckoning in relation to  
 {\em his\/}
plane of simultaneity. The argument that leads to superluminal signals
is sufficiently general that the moving observer will also expect these
to exist, but now in relation to his plane of simultaneity, and so he
would expect the onset of the signal along the second observer's
world-line to be significantly different from what was determined
before. Since onset is an uncontestable physical fact,  this is a
contradiction. The sheer generality of the results leads us to seek a
more fundamental viewpoint from which lack of superluminal communication
would be a consequence and not a hypothesis, much as one supplants
thermodynamics with statistical mechanics and derives the second law
from more basic principles.

The above problem arises because of the dubious mixture of special
relativity with self-subsisting physical states that undergo change
in measurement situations.
Consider a measurement with space-like separated instrumental
events such as a correlation measurement of
the EPR type. In one frame the measurements on the two parts are
simultaneous and so can be considered as just parts of a single
measurement, while in another frame the two measurements are successive
with intervening time evolution. These two description must be equivalent
and produce the same observable results. Thus relativity imposes
constraints that relate the measurement process to the evolution. These
constraints are structurally unstable and neighboring theories are
almost all inconsistent with relativity.

In another study [\refcite{svet:cover}], we explore the nature of these
constraints in a relativistic quantum logic framework. 
This was already presaged in [\refcite{svet:nh}] where it was found that
the absence of gross quality
superluminal effects can be used as supporting
argument for assuming certain axioms in the foundations of quantum
mechanics thus suggesting that quantum mechanics owes some of its
aspects to space-time structure. 

Without going into the details, the axiomatic approach posits a system
of propositions concerning outcomes of experiments performed in regions
of space-time subject to (beyond some standard quantum-logical
impositions) four crucial ingredients: 1) Lorentz covariance, 2)
state transformation due to measurements, taking pure states into pure
states, 3) causality, and 4) something called 
``covariance of objectivity". The second
ingredient is an appropriate generalization of the projection
postulate, the so called ``collapse of the wave function". Depending
only on the state and the measurement arrangement, it incorporates the
basic idea of self-subsisting physical states as interpolating entities
used in calculating joint probabilities of experimental outcomes. 
The third ingredient posits that experimental arrangements in space-like
separated regions are compatible in the technical sense of quantum logic, 
a generalization
of the commutativity of observables in standard quantum theory.
The
fourth ingredient is a technical
 elaboration of Lorentz covariance that is needed
due to the presence of the self-subsisting physical states, as these are
frame dependent entities (they interpolate measurement events in a
temporal sequence, which can be frame dependent). What the postulate
basically means is that if one observer identifies a mixed state as
arising from a measurement process in his causal past
with unknown outcomes, and attributes
to it a decomposition into pure components on the basis of objective
correlations, then another such observer would make the same attributions
using the appropriate Lorentz transformed objects. 

What results from this analysis is that the joint probabilities of
outcomes from space-like separated experiments must satisfy an explicit
constraint. This constraint already precludes the use of long range
correlations for superluminal communications, along the same line of
reasoning as in standard quantum mechanics, but this not too surprising
as one has strong causality ingredients in the axioms. 
What is more interesting is
that if these constraints were to be extended to measurement situation
which are no longer space-like but which are still performed with compatible
instruments, then
one could deduce the famous ``covering law" of Piron's axiomatic quantum
theory [\refcite{piron}], from which a Hilbert space model (not necessarily 
with a complex base field) for the proposition
system follows. 
 
It seems at first hand that there is no way to bridge the 
barrier between the
space-like and time-like compatible arrangements. The presence of 
enough long-range correlations however can do it. Suppose you want to
study right-hand circularly polarized photons. One way is to simply put
an appropriate filter in front of a light source and those photons that
get through are of the right kind and so can be observed at will. Another
equivalent way is to set up an EPR-type arrangement that creates singlet
two-photon states
with the individual photons flying off in opposite directions. 
Put now the
same filter on the {\em distant\/} arm of the EPR apparatus and {\em
nothing\/} on the near arm. Observe at will. Half of the photons
observed are right-hand circularly polarized and half are in the
orthogonal left-hand circularly polarized state, and as the measurements
are done, there is no way of knowing which is which. If all one wants
however is analysis of experimental outcomes, this is no problem, just
wait enough time that the results (passage through the filter or not) 
at the distant arm of each photon
pair are available (typical correlation experiment situation) 
and simply throw out all the experimental data for
the instances where the distant photon did not pass through the filter. 
This provides you with data now  of just the right-hand circularly
polarized photons at the near arm. The fact that these two experimental
procedures are equivalent is a feature of ordinary quantum mechanics
and depends on the existence of a particular entangled state, the
two-photon singlet. 

In the general axiomatic analysis, if one postulates an analogous
equivalence principle, that to any time-like experimental arrangement
with compatible instruments, there is an equivalent space-like
arrangement performed on an appropriate long-range correlated states,
then one completes the argument toward the covering law and a Hilbert
space model of quantum mechanics.

Instead of being simple inconsequential curiosities, as some have
maintained, long-range correlations may be instrumental in making
physics what it is. They provide the link between space-time structure
and mechanics and a bridge between the superluminal and the subluminal. 
Why such a bridge 
should exist cannot be answered at this level of analysis. A more
appropriate scenario would probably be quantum gravity, where the light
cone, and consequently the distinction between superluminal and
subluminal, are emergent concepts and don't exist at the fundamental
level. Apparently the physically relevant solutions for our universe
are such that 
gross quality superluminal situations are suppressed. This should
emerge as a feature of such a theory and not a fundamental ingredient, 
much as quark confinement is a feature of certain gauge theories.

\section{Conclusions}

What can be conclude from all of the above considerations? In the first
place, it's remarkable, as was mentioned in the beginning,  that
apparent superluminal effects have been pointed out in such a wide
variety of theories that ostensibly are relativistic and causal. 
This cannot be
a coincidence and some general characteristic must be at work. A theory
of any complexity about space-time situations may just easily contain
logical implications between propositions concerning situations in 
space-like separated regions, which then may be {\em perceived\/} as
having to do with gross quality superluminal effects. A superluminally
propagating classical solution of Maxwell's equations [\refcite{rod}], 
certainly seems to be a harbinger of such effects.
These perceptions are clearly part of what is 
happening, but it probably is not the full story, and the situation 
certainly bears further study.

If gross quality superluminal effects are found experimentally,
this most likely would radically transform our ideas about the world. 
Experiments should of course be performed, but the question 
then is where to search for these effects. 
The above mentioned apparent superluminal situations in existing
theories seems a natural start, but the situation seems so general that
it's hard to imagine that some of these are just apparent and others
truly lead to gross quality situations. If all are capable of producing
gross quality effects, it's strange that no 
irrefutable experimental evidence
has up to now been forthcoming. It's also unlikely that causal physics is a 
mathematical inconsistency.   The sheer generality of the situations
argues against them.  In the end it seem likely that all these effects
are apparent and any experimental  
attempt based on them to be frustrated.

Another  conclusions  is that 
action-at-a-distance  is a ``soft", that is, a structurally
stable concept. In any formalization  of the space of all
theories it would be characterized by a set of
inequalities (the presence of a non-zero effect)
which would be maintained by any small change in the 
theory. As such it's present in almost any theory one can devise.
By the same token, perception of its possibility in almost any
type of theory should be widespread.
Its absence is a structurally unstable concept.
Fundamental theories that are to be taken seriously should be 
structurally unstable in relation to its fundamental characterizing
properties. Otherwise there would not be
sufficient reason to distinguish them from any neighboring theory. 
Weinberg [\refcite{wein}] argues repeatedly and eloquently 
for the importance of theory rigidity and in this we agree with him.

Taking this into account, advocacy of action-at-a-distance, 
is {\em per se\/} basically counterproductive. It does
not point us to a new fundamental theory. It may be that a new
fundamental theory that supplants the present one 
would have action-at-a-distance  as one of its 
features, but the
new theory would not be characterized by this, 
but by a new rigid set of
properties.

\nonumsection{References}


\begin{thebibliography}{xx}

\bibitem{einst} Einstein,~A., Podolsky,~B. and Rosen,~N.,  {\em
Physical Review\/}, {\bf 47}, 777 (1935).

\bibitem{bohr} Bohr,~N.   {\em
Physical Review\/}, {\bf 48}, 696 (1935).

\bibitem{desp} d'Espagnat,~B. {\em Foundations of Physics}, {\bf 11}, 205
(1981).

\bibitem{rod} Rodrigues,~W.~A.~Jr. and Maiorino,~J.~E. physics/9710030
and references therein.

\bibitem{konst} Konstantinov,~M.~Yu. gr-qc/9810019 and references 
therein.

\bibitem{heger} Hegerfeldt,~G.~C. quant-ph/9809030 and references
therein.



\bibitem{herb} Herbert,~N.,  {\em Foundations of Physics \/},
{\bf 12}, 1171 (1982).

\bibitem{dieks} Dieks,~D.,  {\em
Physics Letters A\/}, {\bf 92}, 271 (1982).

\bibitem{milo} Milonni,~P.~W. and Hardies,~M.~L.,  {\em
Physics Letters A\/}, {\bf 92}, 321 (1982).

\bibitem{woot} Wootters,~W.~K. and Zurek,~W.~H.,  {\em
Nature\/}, {\bf 299}, 802 (1982).

\bibitem{eb2} Eberhard,~P.~H.,  {\em Nuovo Cimento B\/}, {\bf
38}, 75 (1977).

\bibitem{eb3} Eberhard,~P.~H.,  {\em  Nuovo Cimento B\/}, {\bf
46}, 392 (1978).

\bibitem{ghio} Ghirardi,~G.~C.  and Weber,~T.,  {\em
Lettere Nuovo Cimento\/}, {\bf 26}, 599 (1979).

\bibitem{ghim} Ghirardi,~G.~C.,  Rimini,~A.  and
Weber,~T.,
{\em Lettere Nuovo Cimento /}, {\bf 27}, 293 (1980).


\bibitem{foul} There is a vast literature concerning this, which space 
does not allow us to cite here. 
See for instance Foulis,~D.~J. {\em Journal of Natural Geometry\/}, 
{\bf 13}, 1, (1998) and references therein.

\bibitem{wein}Weinberg,~S., {\em Dreams of a Final Theory\/},
Vintage Books (1992). See pp. 88--89 for the discussion on the 
failure of relativistic nonlinear quantum theory.

\bibitem{svet:nlqm} Svetlichny,~G. ``On Relativistic Non-linear 
Quantum Mechanics" in  M.~Shkil, 
A.~Nikitin, V.~Boyko, eds,
{\em Proceedings of the Second International Conference
``Symmetry in Nonlinear Mathematical Physics.
Memorial Prof. W. Fushchych Conference"\/}, Institute of Mathematics of
the National Academy of Sciences of Ukraine, Kyiv, 1997.

\bibitem{hartle} Hartle,~J.~B., ``Spacetime Quantum Mechanics and the
Quantum Mechanics of
Spacetime", in
{\em 1992 Les Houches Ecole d'\'et\'e, Gravitation et
Quantifications\/}

\bibitem{omnes} Omn\'es,~R., {\em The Interpretation of Quantum Mechanics},
Princeton University Press, (1994)


\bibitem{svet:nh} Svetlichny,~G.,  {\em Foundations of Physics\/}, 
{\bf 28}, 131, (1998).

\bibitem{aha1} Aharonov,~Y. and Albert,~D.,  {\em Physical
Review D\/}, {\bf 24}, 359 (1981).

\bibitem{aha2} Aharonov,~Y. and Albert,~D.,  {\em
Physical Review D\/}, {\bf 29}, 228 (1984).


\bibitem{gisin} Gisin,~N.,  {\em Physics Letters A\/}, {\bf
143}, 1 (1990).

\bibitem{elitzur} Elitzur,~A.~C., {\em Physics Letters A\/}, {\bf
167}, 335 (1992).

\bibitem{svet:cover} Svetlichny,~G. ``Lorentz Covariance and the
Covering Law",  pre-print MAT.15/95, Mathematics Department,
Pontif\'{\i}icia Universidade Cat\'olica of Rio de
Janeiro. 
To appear in a revised version in 1999. 

\bibitem{piron} Piron,~C.,  {\em Foundations of Quantum
Physics\/}, W. A. Benjamin, Inc., London (1976).


\end{thebibliography}
\end{document}